\documentclass[10pt]{article}
\usepackage{graphicx}

\def\Title#1{\begin{center} {\Large #1 } \end{center}}
\def\Author#1{\begin{center}{ \sc #1} \end{center}}
\def\Address#1{\begin{center}{ \it #1} \end{center}}

\newcommand\pubblock{\rightline{\begin{tabular}{l} Proceedings of the Fifth Annual LHCP\\ \pubnumber\\
         \pubdate  \end{tabular}}}

\newenvironment{Abstract}{\begin{quotation} \begin{center} 
             \large ABSTRACT \end{center}\bigskip 
      \begin{center}\begin{large}}{\end{large}\end{center} \end{quotation}}

\newenvironment{Presented}{\begin{quotation} \begin{center} 
             PRESENTED AT\end{center}\bigskip 
      \begin{center}\begin{large}}{\end{large}\end{center} \end{quotation}}

\def\Acknowledgements{\bigskip  \bigskip \begin{center} \begin{large}
             \bf ACKNOWLEDGEMENTS \end{large}\end{center}}

\def\beq{\begin{equation}}
\def\eeq#1{\label{#1}\end{equation}}
\def\eeqn{\end{equation}}

\def\beqa{\begin{eqnarray}}
\def\eeqa#1{\label{#1}\end{eqnarray}}
\def\eeqan{\end{eqnarray}}

\let\bar=\overbar

\def\Dslash{\not{\hbox{\kern-4pt $D$}}}
\def\dslash{\not{\hbox{\kern-2pt $\del$}}}

\def\msb{{\bar{\ssstyle M \kern -1pt S}}}

\usepackage{color}

\newcommand\pubnumber{ ATL-TILECAL-PROC-2017-012 }

\newcommand\pubdate{\today}

\def\affiliation{
On behalf of the ATLAS Tile Calorimeter System, \\
Institute of Particle and Nuclear Physics \\
Charles University, Prague, Czech Republic}

\begin{document}

\large
\begin{titlepage}
\pubblock

\vfill
\Title{  Upgrade of the ATLAS hadronic Tile calorimeter for the High luminosity LHC }
\vfill

\Author{ Michaela Mlynarikova  }
\Address{\affiliation}
\vfill
\begin{Abstract}

The Tile Calorimeter (TileCal) is the hadronic calorimeter covering the central region of the ATLAS detector at the LHC. It is a sampling calorimeter consisting of alternating thin steel plates and scintillating tiles. Wavelength shifting fibers coupled to the tiles collect the produced light and are read out by photomultiplier tubes. Currently, an analog sum of the processed signal of several photomultipliers serves as input to the first level of trigger. Photomultiplier signals are then digitized and stored on detector and are only transferred off detector once the first trigger acceptance has been confirmed.

The Large Hadron Collider (LHC) has envisaged a series of upgrades towards a High Luminosity LHC (HL-LHC) delivering five times the LHC nominal instantaneous luminosity. The ATLAS Phase II upgrade, in 2024, will accommodate the detector and data acquisition system for the HL-LHC. In particular, TileCal will undergo a major replacement of its on- and off-detector electronics. All signals will be digitized and then transferred directly to the off-detector electronics, where the signals will be reconstructed, stored, and sent to the first level of trigger at a rate of 40 MHz. This will provide better precision of the calorimeter signals used by the trigger system and will allow the development of more complex trigger algorithms. Changes to the electronics will also contribute to the reliability and redundancy of the system.

Three different front-end options are presently being investigated for the upgrade. Results of extensive laboratory tests and with beams of the three options will be presented, as well as the latest results on the development of the power distribution and the off-detector electronics.

\end{Abstract}
\vfill

\begin{Presented}
The Fifth Annual Conference\\
 on Large Hadron Collider Physics \\
Shanghai Jiao Tong University, Shanghai, China\\ 
May 15-20, 2017
\end{Presented}
\vfill
\end{titlepage}
\def\thefootnote{\fnsymbol{footnote}}
\setcounter{footnote}{0}
%

\normalsize 


\section{Introduction}

ATLAS Tile calorimeter (TileCal) is the central hadronic calorimeter of the ATLAS experiment at the Large Hadron Collider (LHC) \cite{1}. It is a sampling detector with steel plates as absorber and scintillating tiles as active medium. TileCal is divided into the central barrel ($|\eta|<1.0$) and two extended barrels ($0.8<|\eta|<1.7$) along the beam axis. Azimuthally each barrel is segmented into 64 modules, in which scintillating tiles are grouped into individual cells. Each cell is read out by two photomultiplier tubes (PMTs), which are together with on-detector electronics located at the outermost part of each module. 
A series of upgrades is planned for the LHC which will result in the High Luminosity LHC (HL-LHC) with average luminosity 5-10~times larger than nominal instantaneous luminosity of the LHC. 
Large luminosity presents a significant challenge to the detector and trigger and data acquisition system in the form of increased trigger rates and detector occupancy.
Thus in order to accommodate to the HL-LHC conditions, TileCal will undergo an upgrade during which the whole readout electronics will be replaced.

\section{Present readout architecture}
\begin{figure}[h]
\centering
\begin{minipage}[c]{0.85\textwidth}
\includegraphics[width=\textwidth]{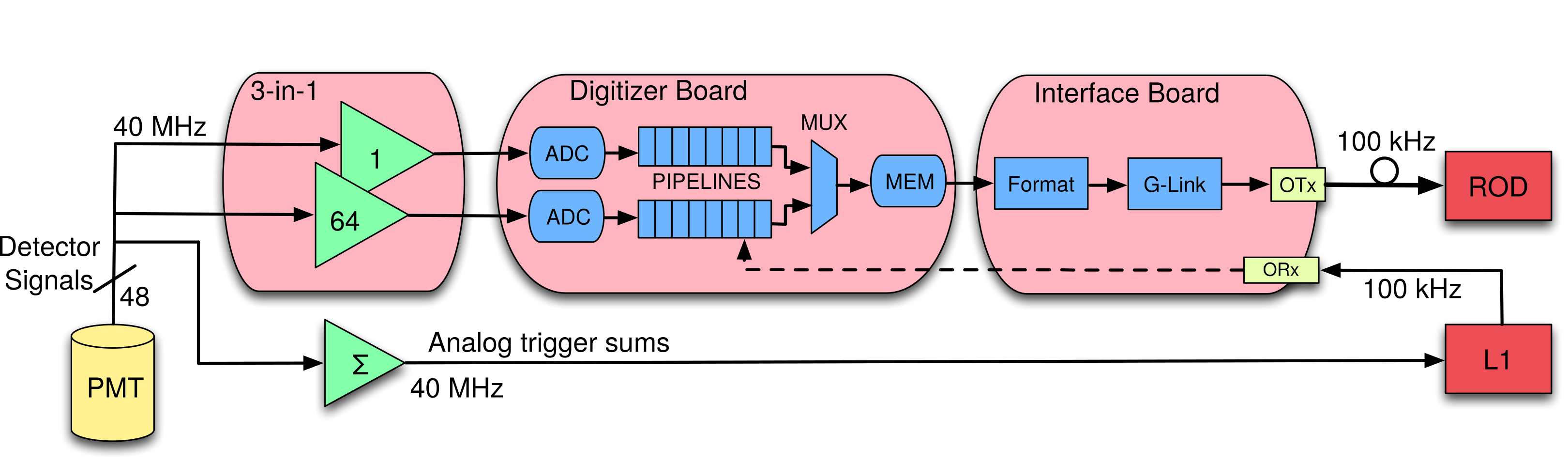}
\end{minipage}
\caption{Sketch of the TileCal front-end readout electronics in the current system. Figure taken from \cite{ele}.}
\label{fig:figure1}
\end{figure}

In the current TileCal signal processing scheme (Fig. \ref{fig:figure1}) the analog pulses from PMTs are shaped and amplified in two readout channels (low and high gain, with a gain ratio of 1:64). Amplified signals from both gains are then sampled and digitized with 10-bit analog-to-digital converters (ADCs) every 25 ns \cite{3}. These digital samples are stored in pipeline memory waiting the Level 1 trigger decision. Simultaneously, the PMT analog signals are grouped and transmitted to the Level 1 Calorimeter system. After receiving the Level 1 trigger acceptance, the digital samples of the selected events are transmitted at a maximum average rate of 100~kHz to the Read Out Driver (ROD) located in the back-end system. The energy, time and quality factors are then reconstructed in RODs for each channel and used in the high level trigger decision chain as well as for the offline data processing. 

\begin{figure}[h]
\centering
\begin{minipage}[c]{0.85\textwidth}
\includegraphics[width=\textwidth]{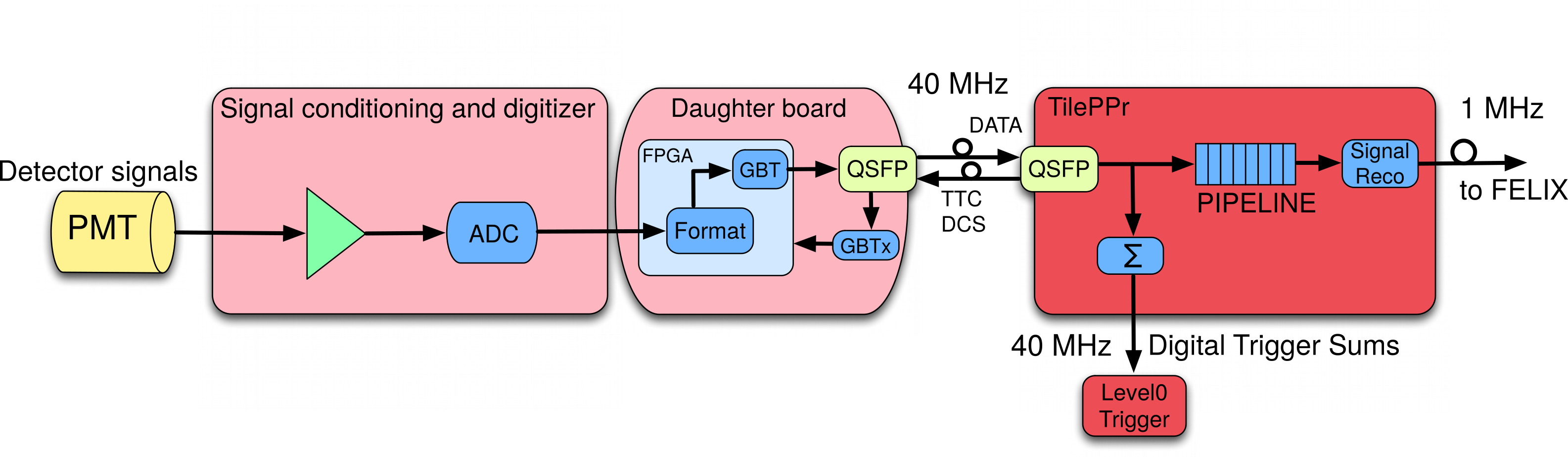}
\end{minipage}
\caption{Sketch of the TileCal Phase-II upgrade readout electronics. Figure taken from \cite{ele}.}
\label{fig:figure2}
\end{figure}

\section{Upgraded readout architecture}
In the upgraded TileCal readout architecture, the current drawer will be divided into four operationally independent units called minidrawers, which will reduce the number of connections and single-point failures in the power and data readout. Each minidrawer will contain 12 PMTs, each with its own Front-End Board (FEBs), a Main Board (MB), Daughter Board (DB) and High Voltage board. 
For every bunch crossing, all data generated in the detector will be transferred to off-detector PreProcessors (TilePPr) \cite{4}, which will provide preprocessed digital trigger information to the ATLAS Level 0 trigger. 
TilePPr will thus represent the interface between the data acquisition, trigger and control systems, and the on-detector electronics. 
The scheme of the new electronic chain is displayed in Fig. \ref{fig:figure2}.

In case of FEB, three different options are being developed and currently undergoing testing:
\begin{itemize}
\item The {3-in-1} option \cite{2} is an improved version of the current front-end board. It is based on discrete components and provides better linearity and lower noise. The FEB includes a passive shaper, bigain amplifiers and a slow integrator. The signals from 12 PMTs are collected in the Main Board, which accommodates 24 12-bit ADCs to digitize the low and high gain (with a ratio of 1:32) of each FEB and four control FPGAs. The data are then transferred to the common DB. 
\item The \emph{QIE} (Charge Integrator and Encoder) is a custom Applications Specific Integrated Circuit (ASIC) that includes current splitter in four gains followed by charge integrator and time-to-digital converter at 40~MHz. 
It does not perform pulse shaping and therefore minimizes pile-up problems and allows to measure raw PMT pulses. 
This ASIC provides digital information which means that the corresponding MB is only responsible for collecting and routing the signals to the common DB. 
\item The \emph{FATALIC} is based on an ASIC chip and includes current conveyor followed by a current integrator with three different gains ($1\times$, $8\times$, $64\times$). The signals from the integrator are digitized with 12-bit ADCs included into the ASIC. The digitized data are then collected in the MB, which subsequently transfers the data to the common DB. 
\end{itemize}

All three front-end options share the same DB, which is responsible for collecting the FEB data and their transmission to TilePPr.
It also receives and decodes configuration commands from the preprocessor and forwards them to the corresponding front-end element. 

The Low Voltage Power Supplies (LVPS) are controlled through dedicated links from the back-end. 

Two different options are being evaluated for the regulation of the High Voltage (HV) needed by the PMTs: 
\begin{itemize}
\item The \emph{remote HV} has the control and monitoring circuitry installed off-detector. This requires an individual HV cable from the counting room to each PMT. This architecture significantly improves the radiation tolerance issues since the electronics are installed off-detector. 
\item The \emph{HV Opto} solution has bulk high voltage power delivered to each front-end drawer and the control and monitoring is implemented in the on-detector electronics to serve the 48 PMTs. In this case the number of long high voltage cables is reduced significantly, but the radiation tolerance issues are tighter. 
\end{itemize}

\section{The TileCal Phase-II Demonstrator and test-beam program}

\begin{figure}[h]
\centering
\begin{minipage}[c]{0.45\textwidth}
\includegraphics[width=0.83\textwidth]{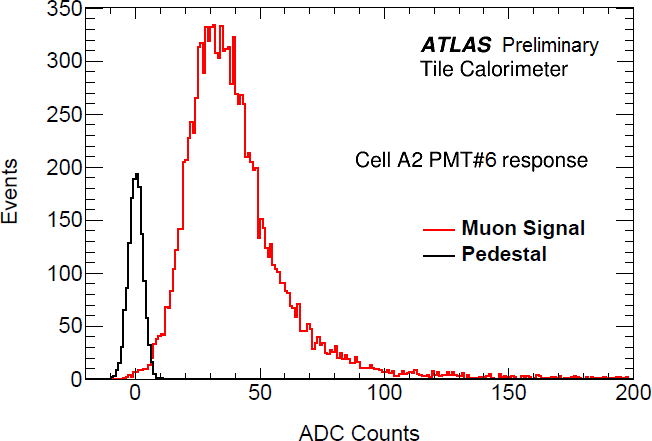}
\end{minipage}
\hspace{5mm}
\begin{minipage}[c]{0.45\textwidth}
\includegraphics[width=0.83\textwidth]{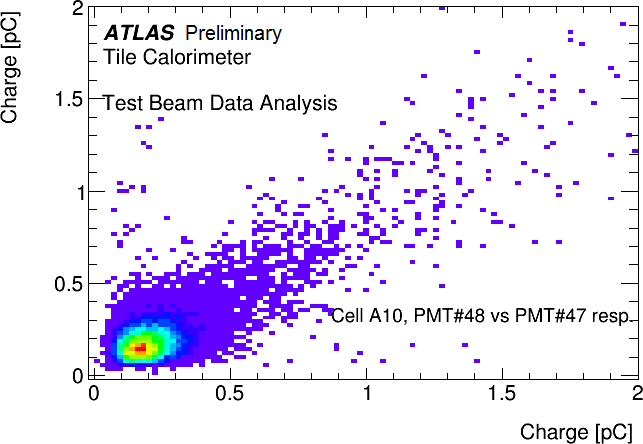}
\end{minipage}
\caption{Left: Muon signal in cell A2 using 100~GeV muons hitting Demonstrator at $20^{\circ}$. Right: Muon signal in cell A10, whose scintillation light is collected from each side by two PMTs ($\#47$ and $\#48$), using 150~GeV muons hitting Demonstrator at $-90^{\circ}$. Figures taken from \cite{Tile}.}
\label{fig:figure3}
\end{figure}

A hybrid TileCal Demonstrator prototype has been developed to evaluate and qualify the new readout and trigger concepts for the full ATLAS operation and data taking. The hybrid Demonstrator uses 3-in-1 front-end option and is fully compatible with the current system providing both the analog signals for the current L1 trigger and the fully digital information for the Phase-II trigger TilePPr prototypes and can replace one of the drawers in the current system. The plan is to install the TileCal Demonstrator in ATLAS during one of the short LHC shutdowns during Run 2 or eventually during the second long shutdown scheduled for 2019. 

Minidrawer prototypes equipped with 3 front-end and 2 HV regulation options were produced to evaluate and compare their operability and performance in standalone tests. 

These prototypes were instrumented during the beam tests with both legacy and Phase-II upgrade electronics and 
exposed to particle beams of various compositions (muons, positrons and hadrons) and energies, in order to evaluate the performance of the detector and the new electronics. Test beam results using muon beam hitting Demonstrator are shown on Figure \ref{fig:figure3}.


\section{Conclusions}
Complete replacement of the TileCal readout electronics is expected for the HL-LHC upgrade. Intense tests and revisions of several components are ongoing to improve their performance.
The front-end electronics prototype has been built and tested in laboratory as well as with the
particle beams in 2015, 2016 and 2017. One more test beam campaign is scheduled for autumn 2017. The FEB and HV option final
selection process have been defined based on extensive comparison of the performances and other
criteria. The production of new electronic modules can start immediately after final selection is
made.


\Acknowledgements
The support from Ministry of Education, Youth and Sports of the Czech Republic is acknowledged.

\end{document}